\documentclass[11pt]{article}
\usepackage{amssymb,amsmath}
\usepackage{slashed}
\usepackage{mathtools}
\usepackage{feynmf}

\usepackage{mathrsfs}
\usepackage{graphicx}
\usepackage{simplewick}
\usepackage{simpler-wick}
\usepackage{amsxtra}

\usepackage{footnote}
\usepackage{tablefootnote}
\usetikzlibrary{graphs,quotes,arrows.meta}

\usepackage[export]{adjustbox}
\usepackage[colorinlistoftodos]{todonotes}
\usepackage[colorlinks=true, allcolors=blue]{hyperref}
\usepackage{hyperref}
\usepackage{authblk}
\DeclareMathOperator{\Lagr}{\mathscr{L}}
\usepackage{tikz}
\usetikzlibrary{graphs}
\usepackage{subfigure}
\usepackage{indentfirst}
\usepackage{cite}
\usepackage{gensymb}
\usetikzlibrary{shapes.geometric, arrows}

\graphicspath{{Pics/}}

\title{Analysis of programming tools in framework of dark matter physics and concept of new MC-generator}
\author{K.M. Belotsky\\ k-belotsky@yandex.ru\\A.H. Kamaletdinov\\kamaletdinov.a.h@yandex.ru\\E.S. Shlepkina\\shlepkinaes@gmail.com}
\affil{National Research Nuclear University MEPhI (Moscow Engineering Physics Institute), 115409, Kashirskoe shosse 31, Moscow, Russia}

\begin{document}
\maketitle

\begin{abstract}
We analyse here some programming tools (MC-generators) from viewpoint of their application to the tasks of dark matter (DM) interpretation of cosmic rays puzzles. We shortly describe our tasks, where the main goal is the solution of the problem of suppression of gamma-rays induced by the products of DM decay or annihilation in Galaxy. We show that existing MC-generators do not fully satisfy our task, comparing them, and suggest our own one.
\end{abstract}


\section{Introduction}\label{s:intro}
The necessity of the usage of different MC-programs\footnote{MC is decoded as Monte Carlo. Such programs are called as ME (Matrix Element) as well, implying the program tools able to simulate new (high energy) physics process.} appears in different areas. One of them is connected with dark matter (DM) processes. DM can give signal in cosmic rays (CR) due to their decay or annihilation. Positron anomaly \cite{Adriani:2008zr, PhysRevLett.113.121101}  or possible excess of electrons and positrons \cite{Ambrosi:2017wek} at high energy in CR is one of such subject.

DM physics is unknown, what requires a respective flexibility of calculations of the predicted signal in CR $e^+e^-$. Realization of this with the help of using some programming tools imposes definite requirements on them about which we will talk. We 
do not pretend to comprehensive review, we are reviewing it from point of view of our task, what can be useful for many adjacent ones too.

The physical task itself comes from our previous works \cite{Belotsky:2016tja,belotsky2019indirect,2019PDU....2600333B,ICPPA-2019,Belotsky:2018vyt,Alekseev2017R, Belotsky:2017wgi,Alekseev2017An,1742-6596-675-1-012026,1742-6596-675-1-012023, Alekseev2015relieving} studying compatibility of DM interpretation of CR $e^+e^-$ with cosmic gamma-ray data. The main problem is that, when we are trying to explain CR $e^+e^-$ anomalies we start to contradict to cosmic gamma-ray data even in the framework of, seeming, minimal model case from viewpoint of gamma-ray production. The latter is pure $e^+e^-$ decay or annihilation mode where gamma appears as (a) FSR (Final State Radiation) and (b) due to interaction of $e^+e^-$ with interstellar medium. Both contributions seem to be unavoidable. Nonetheless, even in this minimal case we got contradiction with gamma-ray data.

There are a few attempts to try to avoid this contradiction (we reviewed them in \cite{belotsky2019indirect,1742-6596-675-1-012026,Alekseev2017An}), i.e.\ to suppress gamma coming from DM. It can relate to specifics of space distribution of DM like clumping or existence of dark disk component (supposing that a dominant halo DM component does not produce CR), or specifics in DM interaction. The latter includes both different decay/annihilation modes and Lagrangian of DM particles interaction with ordinary matter. 

Specifics of DM physics may involve also opportunity of decay of DM particles onto two identical fermions like $X^
{++}\rightarrow e^+e^+$. In such model it is supposed there exist two types of double charged DM particles, $X^{++}$ and $Y^{--}$. It is assumed that the last one is in form of electrically neutral bound state states with He, $X^{++}$ form bound state with $Y^{--}$ and decay \cite{Khlopov2006, belotsky2014dark,doi:10.1142/S0217751X14430027,Belotsky:2014haa}. In case of $X^{++}\rightarrow e^+e^+\gamma$ decay, we have factor two of suppression of FSR gamma per one $e^+$ (because they are two in one decay), and also extra suppression is expected due to identity itself of fermions in final state. The last reason takes place explicitly in classical case (dipole radiation of same charged particles is zero) and
somehow partially in quantum case -- due to so called single photon theorem \cite{brown1995understanding}.

All this accounts for necessity to have respective programming tool able to calculate the processes in the aforementioned tasks and, of course, not only. It does not cancel a desirability of analytical calculations. But the latter is often difficult to do and a crosscheck is necessary even when it is possible. It, in its turn, requires opportunity of step by step tracking calculations making with programming tools.

We demonstrate here the work of some such tools (Section \ref{tools}). They does not provide identical and, therefore, reliable results at the absolutely same initially set parameters. It related to our tasks. We here come to conclusion of creation of MC (HEP) generator (Section \ref{newGenerator},\ref{application}) which would allow simple step by step checking of calculation procedure.

\section{Programming tools analysis}\label{tools}

As we told, it is impossible to build a model of dark matter in framework of dark halo or dark disk that would completely explain the positron anomaly in cosmic rays. Such attempts will lead to an excess of FSR arising from the decay/annihilation of a dark matter particle into two charged leptons or during the propagation in the interstellar medium.

This task requires to create a new physical models that go beyond the Standard Model (BSM).
It is necessary to find the most suitable programming tools for such a task that would correspond the following minimum requirements:
\begin{enumerate}
\item the possibility to implement new physical models (BSM),
\item compute a matrix element and squared matrix element in analytical form,
\item the possibility of an explicit description of charge conjugation,
\item high enough precision of calculation.
\end{enumerate}

To describe the decay or annihilation of DM particles, taking into account possible FSR, the different programming tools such as MadGraph \cite{maltoni2003madevent}, CompHEP \cite{boos2004comphep}, CalcHEP \cite{belyaev2013calchep} and FormCalc \cite{hahn2000automatic} were considered.

Implementing BSM models in a generator such as MadGraph requires describing the model using the FeynRules \cite{christensen2009feynrules} package. FeynRules is a package with Mathematica \cite{wolfram1999mathematica} source code that allows calculating the Feynman rules in momentum space for any physical model of quantum field theory.

One of the reasons for using this package is the possibility of  describing charge conjugation for fermions, which is necessary in our models.

In FeynRules, we started with the following DM models: the
simplest model of DM particle X decay on two opposite charged leptons and the model of double charged scalar particles $X$. In both models particle X is hypothetical long-lived  scalar particle with a mass of about 1-3 TeV. Feynman rules for the Lagrangians presented below, which describes the decay of this particle, were tested:
\begin{equation}
    \Lagr = X\overline{\psi} (a+b\gamma^5)\psi +  \overline{\psi}\gamma^{\mu}A_{\mu}\psi
\end{equation}
\begin{equation}
    \Lagr = X\overline{\psi^C} (a+b\gamma^5)\psi + X^*\overline{\psi} (a-b\gamma^5)\psi^C - \overline{\psi}\gamma^{\mu}A_{\mu}\psi
\end{equation}

where $a$ and $b$ are the unknown constant parameters.

At the output, sets of model files written in the Universal FeynRules Output (UFO) were obtained that can be used for calculations and modeling of various processes in the MC-generator MadGraph5aMC@NLO.

MadGraph is programming tool wich allows calculating cross-sections and squared matrix elements in numerical form.

Using the FeynRules model files, several decay modes of the DM particle X, namely, the processes $X\rightarrow e^+ e^+$ and $X\rightarrow e^+ e^+ \gamma$, were simulated in this generator.
MadGraph allows calculating cross-section, but it does not allow geting the squared matrix element in an analytical form, so this generator does not corresponds to all the previously set requirements.
 
 The next two MC-generators that we used in our task are CompHEP and CalcHEP. These tools have attracted our attention since they have the ability to obtain a squared matrix elements. Obtaining the squared matrix elements in analytical form for each of the processes $X\rightarrow e^+ e^{\pm}$, $X\rightarrow e^+ e^{\pm} \gamma$, we get the opportunity to monitor the correctness of the results and compare them with those that were obtained manually.
 
To implement our models to CalcHEP, one can use the LanHEP \cite{semenov1998lanhep} package. LanHEP has been designed as part of the MC-generator CalcHEP. This package,similar to the  FeynRules package, is used to generate Feynman rules in a momentum representation based on a given Lagrangian.The output can be written in the form of CalcHEP's model files, which allows to start computing processes in a new physical model.

One of the alternatives to the MC-generators that we considered in framework of this task was FormCalc. FormCalc is the tool wich based on the FORM syntax and implemented as Mathematica  package that allows one to calculate Feynman diagrams . Receiving input Feynman diagrams generated by the FeynArts (FeynArts \cite{hahn2001generating} tool for generating Feynman diagrams), FormCalc is able to make calculations of the squared matrix element and write it out in Fortran code. The advantage of this program is that one can see some intermediate results, such as squared matrix element. However, FormCalc is a complex modular system of several packages and tools.

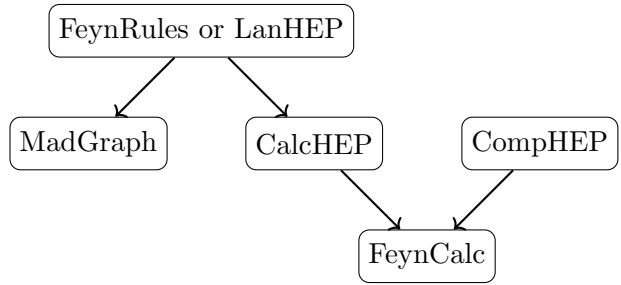
\begin{figure}[h!]
\begin{center}
\begin{tikzpicture}[node distance=1.5 cm]
\tikzstyle{arrow} = [thick,->]
\tikzstyle{block} = [rectangle, rounded corners, minimum width=0.7cm, minimum height=0.7cm,text centered, draw=black]
\node (FR) [block] {FeynRules or LanHEP};
\node (MG) [block, below of=FR, xshift=-1.5 cm] {MadGraph};
\node (CalcH) [block, below of=FR, xshift=1.5 cm] {CalcHEP};
\node (CompH) [block, below of=FR, xshift=4.5 cm] {CompHEP};
\node (Feyn) [block, below of=CalcH, xshift=1.5 cm] {FeynCalc};
 \draw[arrow] (FR) -- (MG);
 \draw[arrow] (FR)-- (CalcH);
  \draw[arrow] (CalcH)-- (Feyn);
    \draw[arrow] (CompH)-- (Feyn);

\end{tikzpicture}
\end{center}
\caption{Approximate schemes for working with some MC-generators}
\label{scheme1}
\end{figure}

\begin{figure}[h!]
\begin{center}
\begin{tikzpicture}[node distance=1.5 cm]
\tikzstyle{arrow} = [thick,->]
\tikzstyle{block} = [rectangle, rounded corners, minimum width=0.7cm, minimum height=0.7cm,text centered, draw=black]
\node (FR) [block] {FeynRules};
\node (FA) [block, below of=FR, xshift=0cm] {FeynArt};
\node (FC) [block, below of=FA, xshift=0cm] {FormCalc};
\node (Fcode) [block, below of=FC, xshift=0cm] {Fortran Code};
\node (dots) [block, below of=Fcode, xshift=0cm] {$\ldots$};
\draw [arrow] (FR) -- (FA) ;
 \draw[arrow] (FA)-- (FC);
 \draw[arrow] (FC)-- (Fcode);
 \draw[arrow] (Fcode) -- (dots);
\end{tikzpicture}
\end{center}
\caption{Modular system of FormCalc using}
\label{scheme2}
\end{figure}
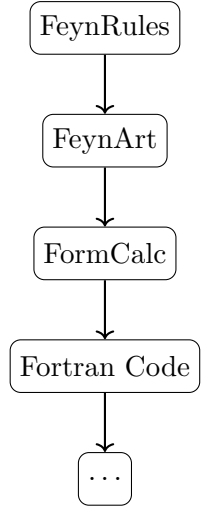

Figures ~\ref{scheme1} and ~\ref{scheme2} show approximate schemes for working with some MC-generators.

The main task at the first stage was the need to determine which programming tools is the most suitable for aforementioned task. 
An analysis of the above MC-generators was carried out, which consisted in comparing the results obtained from different MC-generators using the same model created using LanHEP.
A positive result would be a complete

\begin{figure}[h!]
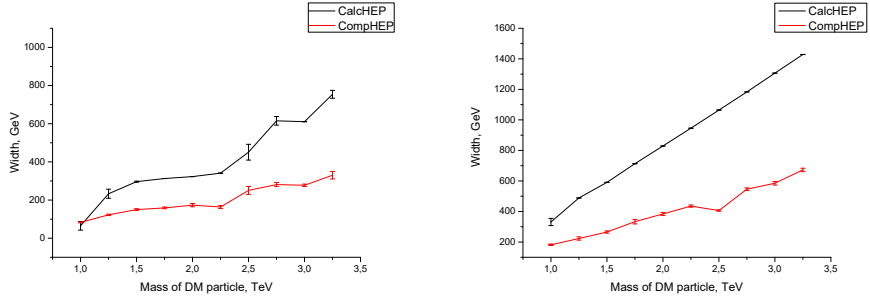

\begin{minipage}[h]{0.49\linewidth}
{\includegraphics[width=0.98\linewidth]{graphe-e-.png}}
\end{minipage}
\begin{minipage}[h]{0.49\linewidth}
\center{\includegraphics[width=0.98\linewidth]{graphe+e-.png} }
\end{minipage}
    \caption{Comparative analysis of MC-generators, using two processes as an example. Left: $X\rightarrow e^+ e^-\gamma$ , Right: $X\rightarrow e^+ e^+\gamma$}.
\label{graph}
\end{figure}

(within the errors) agreement between their results. We considered dependencies of the decay width of the DM particle on its mass (fig.~\ref{graph}). 
These graphs do not show the results obtained from the MadGraph MC-generator, since the decay width obtained using this tool is too large and could not be used in the general analysis. The reason for such deviations has not yet been found. 

Figure ~\ref{graph} shows the results of the tests. As can be seen, the decay widths for the same model and masses of particle $X$
differ.
This deviation motivates us to look for additional verification tools.

It is almost impossible to determine the cause of such differences, since in the process of decay modeling it is impossible to obtain any intermediate results, such as, for example, matrix elements, etc.

The summary table (table  ~\ref{table}) of the capabilities of some MC-generators was compiled, as applied, in particular, to BSM processes.

Summing up, we can conclude that none of the programming tools we have use are not fully suitable for our task.

\begin{table}[h!]
\begin{center}
\begin{minipage}{12cm}
\begin{tabular}{ |p{3.5 cm}|c|c|c|c| }
\hline
Options & CompHEP & CalcHEP & Madgraph & Pythia \\
\hline
Implementing of new models & + & + & + & $-$\\
\hline
Charge conjugation & + & + & + & $-$ \\
\hline
Matrix element in analytical form & $-$ & $-$ & $-$ & $-$ \\
\hline
$|M|^2$ in analytical form & + & + & $-$ & $-$ \\
\hline
 High precision & ${\pm}$\footnote{Hereinafter, the sign ${\pm}$ will mean that this tool does not fit exclusively to our task, but it copes well with other processes and models.} & ${\pm}$ & ${\pm}$ & + \\
 \hline
 Performance\footnote{Characterizes the speed of calculations} & ${\pm}$ & + & ${\pm}$ & ${\pm}$ \\ 
 \hline
 Have an implementation packages\footnote{New models can be loaded into CalcHEP  and MadGraph with the help, for example, FeynRules and LanHEP packages, while in CompHEP one can add new models only by hand. 
 } & $-$ & + & + & $-$\\
 \hline
 Hadronisation & $-$ & $-$ & $-$ & + \\
 \hline
\end{tabular}
\end{minipage}
\end{center}
\caption{Comparison of different MC-generators from viewpoint of calculation DM particle processes.}
\label{table}
\end{table}

\section{Idea of creating of new MC-generator}
\label{newGenerator}

From analysis of existing MC-generators, given above, we come to conclusion that there is so far a necessity of creation of new one adjusted for our (of course, not only) tasks. 
The proposed new HEP generator allows calculating and displaying all intermediate results of calculations - i.e. analytical form of matrix element, the square of the matrix element in the form of traces of gamma matrices, the square of matrix element in form of kinematic variables and result of integrating the square of the matrix element of the given process over the phase volume.

Estimation of intermediate calculation results can be useful for validation of calculation processes and in the phenomenological areas of high energy physics to understand the contribution of specific Lagrangian terms to the various  distributions.

In specific of our work on dark matter interaction physics \cite{Belotsky:2016tja,belotsky2019indirect,2019PDU....2600333B,ICPPA-2019} we need to estimate why given components of Interaction Lagrangian lead to certain effects.

The developing generator is based on FORM symbolic manipulation system \cite{2017arXiv170706453R}, which is designed to work with algebraic expressions and constructions. It reads text files containing definitions of mathematical expressions as well as statements which tell it how to manipulate these expressions. It is widely used in the theoretical particle physics community, but it is not restricted to applications in this specific field.

FORM "doesn't know" anything about the particle physics processes and calculations of amplitudes and cross sections. Everything that FORM makes - it searches in the string the substrings matching the pattern and replaces them with the developer-specified expressions. Then it leads similar terms and displays the result.

User can enter the expression of Lagrangian or the expression of partial term of a perturbation theory series. It is also necessary to explicitly indicate the types of fields used in the Lagrangian and "in" and "out" states. See Figure \ref{structure}.
\begin{center}
\begin{figure}[h!]
\begin{tikzpicture}[node distance=1.6cm]
\tikzstyle{arrow} = [thick,->]
\tikzstyle{user} = [circle, rounded corners, minimum width=1.5cm, minimum height=1.5cm,text centered, draw=black]
\tikzstyle{block} = [rectangle, rounded corners, minimum width=0.7cm, minimum height=0.7cm,text centered, draw=black]
\tikzstyle{data} = [rectangle, dashed, rounded corners, minimum width=0.7cm, minimum height=0.7cm, text centered, draw=black]

\node (user) [user] {User};

\node (fields) [data, below of=user, xshift=2cm] {Fields};
\node (lagrangian) [data, below of=user, xshift=0cm] {Lagrangian};
\node (io) [data, below of=user, xshift=4.6cm] {In and out states};

\node (interface) [block, below of=fields, xshift=0cm] {Interface};

\node (me) [block, below of=interface, xshift=0cm] {Matrix element calculation};

\node (meout) [block, right of=me, xshift=3cm] {Matrix element};

\node (squaring) [block, below of=me, xshift=0cm] {Squaring Matrix Element};

\node (tr) [block, right of=squaring, yshift=0.5cm, xshift=3cm] {$|M|^2 = Tr(...)$};
\node (mom) [block, below of=tr, yshift=0.5cm, xshift=0cm] {$|M|^2 = F(p_{out})$};

\node (output) [block, right of=squaring, xshift=6cm] {Output};

\node (int) [block, below of=squaring, xshift=0cm] {Integral over phase volume};

\node (distr) [block, right of=int, yshift=-0.3cm,xshift=3cm] {Distribution};

\draw [arrow] (user) --node[anchor=east] {Input} (lagrangian) -- (interface);
\draw [arrow] (user) -- (fields)  -- (interface);
\draw [arrow] (user) -- (io)  --
(interface);
\draw [arrow] (interface) -- (me);
\draw [arrow] (me) -- (meout);
\draw [arrow] (me) -- (squaring);
\draw [arrow] (squaring) -- (int);
\draw [arrow] (squaring) -- (tr);
\draw [arrow] (squaring) -- (mom);
\draw [arrow] (meout) -- (output);
\draw [arrow] (tr) -- (output);
\draw [arrow] (mom) -- (output);
\draw [arrow] (int) -- (distr);
\draw [arrow] (distr) -- (output);
\draw [arrow, rounded corners] (output) |- (user);
\end{tikzpicture}
\caption{General structure of the developing generator modules}
\label{structure}
\end{figure}
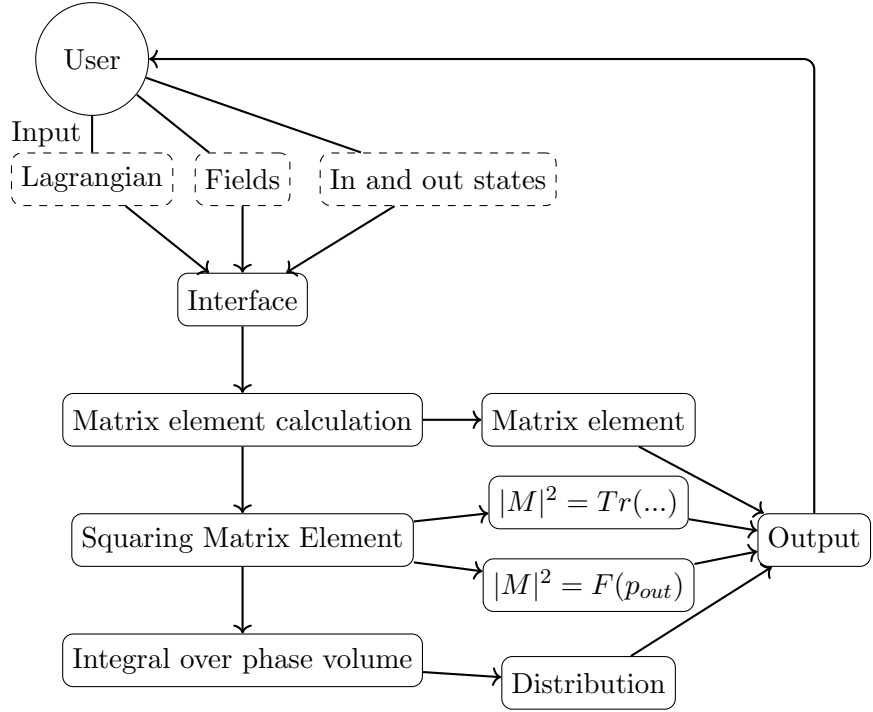
\end{center}
We want to note the monolithic architecture of the developing generator. That is all described above tasks are performed within one single program.
\newline
The matrix element calculation algorithm is based on the principle of secondary canonical quantization. That is if user enter the expression of lagrangian, program approximate the T-exponent by Teylor series, that give the perturbation theory series.

\begin{equation}
    e^{-iS} = 1 - iS + {(-iS)^2 \over 2} + {(-iS)^3 \over 3!} + ...
\end{equation}
where $S \equiv \int d^4x \Lagr$ - is the action of model. 

And take interesting term of this one. After that generator takes the fields of considering model and performs the second quantization\footnote{This means that the symbols $\Phi$ are replaced by other text expressions corresponding to operators.}:

\begin{equation}
\begin{gathered}
    \Lagr \equiv \Lagr(\phi, \partial_{\mu} \phi) \\
    \phi \rightarrow \hat{\phi} \equiv \int {d^3 p \over (2\pi)^3} {1 \over \sqrt{2 \omega_p}} (\hat{a}_p e^{-ipx} + \hat{a}^{\dag}_p e^{ipx})
    \label{secondq}
\end{gathered}
\end{equation}
where $\hat{a}_p$ - is the lattice operator such that $[\hat{a}_p,\hat{a}^{\dag}_q] = (2\pi)^3 \delta^{(3)}(p - q)$

FORM can perform specified instructions with given expressions taking into account the non-commutativity of variables.

Developing generator should include explicitly the permutation rules of the given non-commuting variables in the form of instructions which patterns should be replaced by other expressions.

 That is the replacing of bosonic rising operators at each iteration schematically looks like:
\begin{equation}
    ... \cdot \hat{a}_{\rm p} \hat{a}^{\dag}_{\rm q} \cdot ... \rightarrow ... \cdot \big( (2\pi)^3 \delta^{(3)}(\rm{p} - \rm{q}) - \hat{a}^{\dag}_q \hat{a}_p \big) \cdot ...
    \label{replacing}
\end{equation}

Then program takes the expression of matrix element in form of approximated T-exponent by the Teylor series with second quantization (see Eq.\ref{secondq}):

\begin{equation}
    \Big\langle out \Big| e^{-iS} \Big| in \Big\rangle = \Big\langle out \Big| \Big(1 - iS + {(-iS)^2 \over 2} + {(-iS)^3 \over 3!} + ...\Big) \Big| in \Big\rangle.
\end{equation}
Here $| in \rangle \equiv \hat{a}^{\dag}_{q_1} ... \hat{a}^{\dag}_{q_k} |0 \rangle$ and $\langle out | \equiv \langle 0 | \hat{a}_{p_1} ... \hat{a}_{p_n}$ are the initial and final states of process which are specified by user and are expressed by specific character sets.

Then the program performs the normal ordering of rising operators according to the instructions indicated explicitly in the algorithm and described schematically (\ref{replacing}) above.

One of features of the developing generator is the opportunity for the user to indicate perturbation theory order, as well as choose or enter only the interesting term of perturbation theory for consider only it's contribution.

After the matrix element of the process has been calculated - its analytical expression is displayed to the user on the screen (See Figure \ref{structure} - Matrix element calculation).

The part of the program described above has already been developed.

The next block of the algorithm in the Figure \ref{structure} (Squaring of the matrix element) takes an expression for the matrix element, which was calculated in the previous block of the diagram, and builds an expression for hermitian conjugate operator in the form of a specific string of characters.

Then the product $|M|^2 = M \cdot M^{\dag}$ should be reduced to a trace of gamma matrices and displays to the user.

After substituting kinematic variables into the obtained expression and taking the trace, integration over the phase volume is performing to obtain the distribution.

\section{Application of programming tools}\label{application}
We compare the results, computed by developing generator with the standard processes of particle physics and the specific processes of our work, previously calculated manually.
The results are follows:

1)Two-particle decay of a neutral Dark Matter particle into an electron and a positron
user enter the fields $X$, $\Psi$, $\bar{\Psi}$ and interaction lagrangian of the model
\begin{equation}
    \Lagr = X \bar{\Psi} (a + b \gamma^5) \Psi
\end{equation}
Then he indicates the statistic of fields, that is $X$ - is the scalar field and $\Psi$ - is the spinor field.

This leads to:
\begin{equation}
    M = FB(e,k_1) \cdot (a+b\cdot G(5)) \cdot FC(e,k_2) \cdot S(X,k_3)
\end{equation}
that means:
\begin{equation}
    M = \bar{u}(k_1) (a+b\gamma^5) v(k_2)
\end{equation}

2) Two-parrticle decay of a double charged Dark Matter particle into two positrons.

Similarly:
\begin{equation}
    \Lagr = X\bar{\Psi}(a+b\gamma^5)\Psi^{(c)} + H.C.
\end{equation}
with fixed initial and final states as $\rm |in> \equiv |X>$ and $\rm |fin> \equiv |e^+,e^+>$

\begin{equation}
\begin{gathered}
    M = -FCT(e,k_1)\cdot i G(2)\cdot G(0) \cdot (a+b\cdot G(5)) \cdot FC(e,k_2) \cdot S(X,k_3) + \\ + FCT(e,k_2) \cdot i G(2) \cdot G(0) \cdot (a + b \cdot G(5)) \cdot FC(e,k_1) \cdot S(X,k_3)
\end{gathered}
\end{equation}
that means:
\begin{equation}
    M = - v^T(k_1) i \gamma^2 \gamma^0 (a+ b\gamma^5) v(k_2) + v^T(k_2) i\gamma^2 \gamma^0 (a+b\gamma^5) v(k_1)
\end{equation}

\section{Conclusion}\label{conclusion}

Here we considered capabilities of several MC-generators (CompHEP, CalcHEP, MadGraph with applications to some of them such packages as LanHEP, FeynRules and etc. and some modular tools like FormCalc). This was done in framework of our task concerning DM signal search in CR. More concretely, we considered decay of DM particles with different interaction Lagrangians. We see that the considered tools do not quite satisfy our requests. We need some single tool what would allow providing to show ``step by step'' results of calculations. We suggest it here on the base of code FORM.

\section{Acknowledgements}

Our work 
was supported by the Ministry of Education and Science of the Russian Federation, MEPhI Academic Excellence Project (contract  02.a03.21.0005, 27.08.2013). The work of K.B. is also funded by the Ministry of Education and Science of the Russia, Project  3.6760.2017/BY. 

We want to express special thanks to Lensky Vadim A. for a detailed explanation of the principles of working with FORM.

\bibliographystyle{ieeetr}
\bibliography{main}

\end{document}